\documentclass[prb, aps, twocolumn, groupedaddress,showpacs]{revtex4}
\usepackage{eqnarray}
\usepackage{enumitem}
\usepackage[french,english]{babel} 
\usepackage{amsmath,amsfonts,amssymb}
\usepackage{graphicx}
\usepackage{float}
\usepackage{color}
\usepackage{braket}
\usepackage{version}

\begin{document}

\title{All optical Controlled-NOT quantum gate based on an exciton-polariton circuit}

\author{D. D. Solnyshkov}
\author{O. Bleu}
\author{G. Malpuech}
\affiliation{Institut Pascal, PHOTON-N2, Clermont Universit\'e, Blaise Pascal University, CNRS,24 Avenue des Landais, 63177 Aubi\`ere Cedex, France}

\pacs{71.36.+c 03.75.Lm 14.80.Hv}
\begin{abstract}
We propose an implementation of a CNOT quantum gate for quantum computing based on a patterned microcavity polariton system, which can be manufactured using the modern technological facilities. 
The qubits are encoded in the spin of polaritons. The structure consists of two wire cavities oriented at 45 degrees with a micropillar between them. The polariton spin rotates due to the Longitudinal-Transverse splitting between polarisation eigenstates in the wires. In the pillar, the optically generated circularly polarised polariton macrooccupied state plays the role of the control qubit. Because of the spin-anisotropic polariton interaction, it induces an effective magnetic field along the Z-direction with a sign depending on the qubit value. 
 
\end{abstract}
\maketitle

Quantum computing has evolved a lot since the original idea of R. Feynmann\cite{feynmann}. Several quantum algorithms outperforming their classical analogs have been proposed and implemented more or less successfully. These algorithms, based on the quantum parallelism, target such problems like factoring large numbers into prime numbers \cite{Shor}, optimization\cite{Somma}, and search \cite{PhysRevA.66.042310}. The incredible possibilities offered by quantum computers made the scientists invest a lot of efforts in this field. However, the implementation of these algorithms is haunted by serious obstacles, the most important one being the rapid decoherence of quantum bits (qubits). Various physical implementations of these algorithms have been proposed, the most important difference between them being the choice of the physical realization of the quantum states for encoding the qubits. 
The implementations can be based on discrete quantum states, such as the confined states of the quantum dots \cite{PhysRevA.57.120}, on the spin degree of freedom, as in liquid-state nuclear magnetic resonance \cite{NMR}, or on different combinations of the states of Bose-Einstein condensates \cite{PhysRevA.85.040306}. 

All these degrees of freedom can be combined to encode information if one decides to make use of quantum microcavities in the strong coupling regime, and the corresponding 2-dimensional quasiparticles - exciton-polaritons. These particles are a superposition of light (photons confined in the microcavity) and matter (excitons in the quantum wells) \cite{CavityPolaritons}. They can be easily crated, controlled, and detected using optical means, and their polarization (spin) degree of freedom is easy to manipulate and measure as well \cite{ReviewSpin}. Their in-plane spatial confinement can be organized by patterning the microcavity \cite{PhysRevB.59.10251,Dasbach05,Gao2012,Wertz2010,Galbiati2012,TaneseSolitons} and/or by applying external potentials, which can be created optically \cite{Tanese,Baumberg} or induced by surface acoustic waves \cite{Krizhanovskii}. Finally, polariton Bose-Einstein condensates are also readily available \cite{Kasprzak}, even at room temperature \cite{Baumberg,Feng}.

Using polaritons to implement quantum bits provides many advantages. Polaritons, thanks to their photonic fraction, propagate very rapidly, which allows to reduce the problems with decoherence. Their spin relaxation length exceeds hundreds of microns in recent experiments \cite{Amo2012,Lagoudakis}, while for electrons it is typically several microns \cite{Wang2013}.  
A recent work proposes to use the polariton Rabi oscillations \cite{PhysRevLett.112.196403} as a basis for qubit representation. This approach, however, is limited by the use of the strongly damped upper polariton branch, which leads to rapid decoherence of the qubit \cite{Laussy2014}, and by the difficulties with the control of the qubit state, requiring large energy shifts.
We propose to use the polarization degree of freedom of polaritons to encode information. For example, the circular-polarized $\sigma^+$ state can be assigned a logical 0 ($\left|0\right\rangle$), and the $\sigma^-$ state can be assigned a logical 1 ($\left|1\right\rangle$). A generic qubit is a superposition $\alpha\left|0\right\rangle+\beta\left|1\right\rangle$ corresponding, in general, to the elliptic polarisation of light. By definition, a quantum gate acts simultaneously on both components of the basis (both circular components). In practice, the state of such a qubit can be modified using effective magnetic fields, well known in quantum microcavities \cite{ReviewSpin}. Such fields can be in-plane, physically induced by the energy splitting which exists between the TE and TM optical modes in planar cavities. In a 1D patterned wire cavities\cite{PhysRevLett.97.066402}, this splitting generally lies between the Longitudinal and Transverse modes. Along the $Z$-direction, a real applied magnetic field can act on polaritons by inducing a finite Zeeman splitting of polariton states \cite{Fischer2014,Sturm2014}. A self-induced effective field along the $Z$-direction can also be created due to the polariton-polariton spin anisotropic interaction \cite{anisotropic}. Indeed, the polarization degree of freedom of polaritons has already been proposed as a possible solution for the implementation of classical optical logic gates \cite{PhysRevLett.99.196402}. That a qubit based on polariton spin can be initialized to an arbitrary value and that this value can be maintained for a long time does not need to be proven: such experiments were already carried out \cite{PhysRevLett.109.036404,Amo2012}, although the polariton state has not been considered as a qubit in these works. It was shown that the decoherence time for polaritons is much longer than the lifetime. Here, we demonstrate how the two mechanisms of the qubit control based on effective magnetic fields can be combined together in order to achieve the expected operation of the CNOT (controlled NOT) gate. This gate is the essential quantum gate: any quantum algorithm can be implemented using only the CNOT double-qubit gate and single-qubit rotations. 

This paper is organized as follows. First, we describe the functioning of a CNOT gate and its implementation. Second, we describe the analytical and numerical models used for its simulation. In the last section, we present the results obtained with these models and conclude on the feasibility of the device.

\section{The CNOT gate implementation}

If the polarization basis is used to encode the state of the qubit, the Bloch sphere of the qubit is naturally mapped to the pseudospin vector of light, whose elements can be defined through the components of the spinor the following way:

\begin{eqnarray}
S_{0}&=&\sqrt{S_x^2+S_y^2+S_z^2}\\
{S_{x}} &=&\Re \left( {\phi _{+}\phi _{-}^{\ast }}\right)\nonumber \\
{S_{y}} &=&\Im \left( {\phi _{+}^{\ast }\phi _{-}}\right)\nonumber \\
{S_z} &=& \left( {{{\left| {{\phi _ + }} \right|}^2} - {{\left| {{\phi _ - }} \right|}^2}} \right)/2 \nonumber
\end{eqnarray}

We note that if the pseudospin is normalized to unity, the $S_z$ component is nothing but the degree of circular polarization $\rho_c=(n_{\phi_+}-n_{\phi_-})/(n_{\phi_+}-n_{\phi_-})$ of the emission.

The states $\left|0\right\rangle$ and $\left|1\right\rangle$ of the qubit correspond to the $+Z$ and $-Z$ directions of the pseudospin, while their linear combinations correspond to other points of the pseudospin sphere. For example, the combination $\left( {\left| 0 \right\rangle  + \left| 1 \right\rangle } \right)/\sqrt 2$ corresponds to the $+X$ direction of the pseudospin vector, which is usually defined as corresponding to the horizontal polarization of emitted light.  In general, a qubit is a vector whose endpoint travels on the surface of the pseudospin (or Bloch's) sphere. Because of the decoherence, its length can be decreased, thus representing partially or completely unpolarized light. In our consideration, the decoherence time is much longer than the lifetime, and therefore the partial depolarization can be neglected from the consideration.

\begin{figure}[bp]
\includegraphics[scale=0.40]{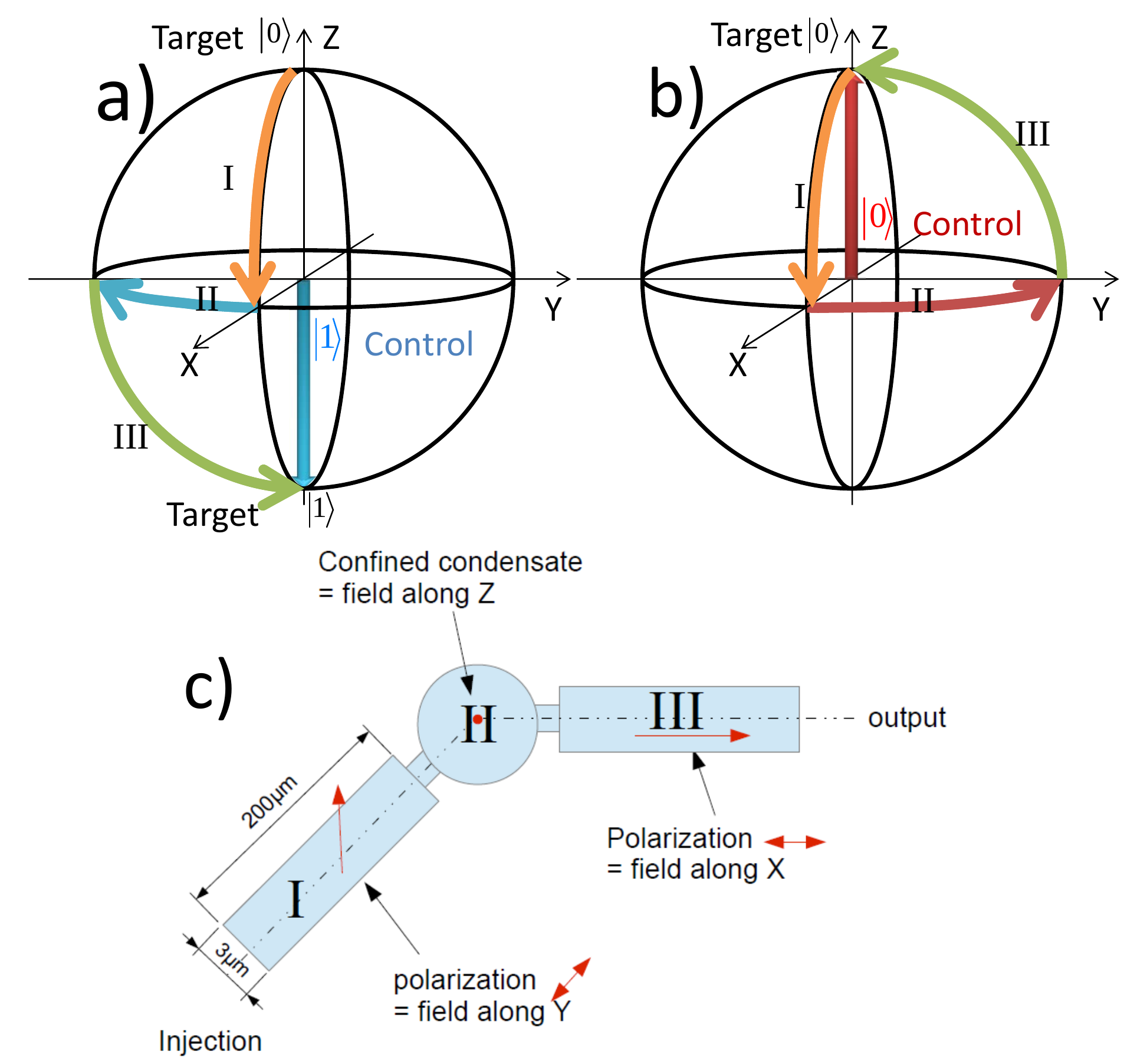}

\caption{(Color online) Operation of the CNOT gate shown on a Bloch sphere. The arrows on the surface of the sphere demonstrate the evolution of the target qubit. Vertical arrows show the control qubit and the corresponding magnetic field. a) Control qubit in the state $\left| 1 \right\rangle$, target qubit inverted; b) Control qubit in the state $\left| 0 \right\rangle$, target qubit not inverted; c) Physical implementation scheme based on wire cavities and a micropillars. Red arrows show the effective fields. The operation steps are marked in roman numbers. }
\label{principe}
\end{figure}

The CNOT gate operates with 2 qubits: the target qubit and the control qubit. Their representation on the Bloch's sphere, and the principle of the operation of the CNOT gate are shown in figure 1 (panels a,b), which  demonstrates the operation of the CNOT gate in the two cases: a) with the control qubit being $\left|1\right\rangle$ (the target qubit should become inverted) and b) with the control qubit being $\left|0\right\rangle$ (the target qubit should not be inverted). In both cases, the target qubit is initially in the state $\left|0\right\rangle$ ($+Z$ direction).

The first step consists in placing the target qubit in the $XY$ plane by using an effective magnetic field along $Y$, which causes the rotation from $+Z$ to $+X$ (step I, orange arrow on the sphere surface). This step should not depend on the state of the control qubit.

The second step involves the effective field created by the control qubit on the target one. Indeed, as a result of the spin-anisotropic polariton-polariton interactions \cite{ReviewSpin}, circularly polarized polaritons create an effective field acting on the polariton pseudo-spin and pointing along the $Z$ axis depending on the circularity. Therefore, as a function of the state of the control qubit, this field is either in $+Z$ (control $\left|0\right\rangle$)or in $-Z$ direction (control $\left|1\right\rangle$), and thus the target qubit rotates either towards the $+Y$ or to $-Y$ direction, as shown by red and blue arrows respectively (step II).

Finally, at the third step, another in-plane field is applied to the target qubit, bringing it back to the $Z$ axis, either to the $\left|0\right\rangle$ or to the $\left|1\right\rangle$ state, depending on its previous position, as shown by the green arrow (step III).

In order to provide consecutive action of several effective magnetic fields without changing the external parameters with time, which is difficult on a picosecond timescale, one has to organize the propagation of the target qubit through the system. At the same time, the control qubit, which has to stay constant, need not be propagating. We shall therefore combine localized and propagating polaritons in a patterned microcavity structure (figure 1c), which can be manufactured using the modern technological facilities \cite{Tanese,HaiSon}. The structure shown in figure 1c) consists of a circular pillar, coupled through potential barriers with two polariton 1D wires (input and output), oriented at well-defined angles. The circular pillar contains the control qubit in its confined state. The operation is performed in steps (marked with roman numbers in all 3 panels):
\begin{enumerate}[label=\Roman*]
\item The target qubit is injected into the input wire, where it experiences the effective magnetic field along $Y$ due to the inherent LT splitting of the wire \cite{Tanese}.
\item It then tunnels through the barrier (narrow part of the wire, with a higher transverse localization energy) and enters the pillar, where its polarization rotates again under the effect of the spin-anisotropic interaction with the control qubit (field along $Z$).
\item Finally, the polaritons forming the target qubit exit the localized area through the barrier and propagate in the output arm under the action of an effective field along $X$, caused by the LT splitting (different from that in the input arm due to its different spatial orientation). 
\end{enumerate}  
One should note that the control qubit does not expand into the wires, because the  energy of the lowest quantized state which it is occupying, lies below the propagative states of the wires.

\section{Model}

Polaritons in a quantum microcavity represent a 2D system, even if the cavity is patterned. Moreover, the most exact description with a correct treatment of polarization can only be obtained by solving the Maxwell's equations in 3D. In order to solve the problem analytically, we have to make several approximations. First of all, the Maxwell's equations can be replaced by the Schrodinger equation in 2D, the mass of photons and polaritons appearing in this equation due to the confinement of the cavity modes in the $Z$ direction. The use of the two spin projections for this Schrodinger equation allows to take into account the two polarizations of photons within the pseudospin formalism. The energy splittings between these polarizations are treated as effective magnetic fields acting on the pseudospin. Finally, the 2D spinor Schrodinger equation is reduced to 1D, with a spatially varying potential corresponding to the wire width \cite{Tanese} and spatially varying effective magnetic fields corresponding to the orientation of the wire\cite{BerryPhase}.

For the analytical treatment, we describe the potential barriers by Dirac's delta functions. Finally, proposing the least complicated, proof of principle configuration for the experiment, we assume that the control qubit is created by a non-resonant spin-polarized pumping inside the localized pillar \cite{Fischer2014}, creating an effective magnetic field in the $Z$ direction. A more complete and realistic description of the gate is then achieved by numerical simulations which takes into account a resonant circularly-polarized pulse creating the control qubit.

Our goal is to solve the spinor Schrodinger equation in 1D:

\begin{equation}
 i\hbar\frac{\partial\Psi}{\partial t}=\left(-\frac{\hbar^2}{2m}\Delta+V(\delta(x)+\delta(x-a))+\mathbf{\Omega}(x)\mathbf{\mathrm{\sigma}}
 \right)\Psi 
 \end{equation}

 where the effective fields are spatially-dependent:

\begin{eqnarray}\Omega_Z(x)=
\left\lbrace
\begin{array}{ccc}
V_0  & \mbox{if} & 0<x<a\\
0 & \mbox{} & \mbox{otherwise}
\end{array}\right.
\end{eqnarray} 
\begin{eqnarray}
\Omega_X(x)=
\left\lbrace
\begin{array}{ccc}
0  & \mbox{if} & x<a\\
U_1 & \mbox{if} & x\geqslant a
\end{array}\right.
\end{eqnarray} 
\begin{eqnarray}
\Omega_Y(x)=
\left\lbrace
\begin{array}{ccc}
U_2  & \mbox{if} & x\leqslant 0\\
0 & \mbox{if} & x>a
\end{array}\right.
\end{eqnarray} 

Here, $m\approx 5\times 10^{-5}m_0$ is the polariton mass ($m_0$ is the free electron mass). The spin-anisotropic interaction with the circular-polarized excitonic reservoir is described by an effective field $V_0$ corresponding to the energy difference of the two spin components between the delta barriers. Thus, as discussed above, the control qubit is replaced by a classical effective field, in order to simplify the analysis in this section. $\Omega_X$ and $\Omega_Y$ are the constants determining the effective field related to the LT polarization splitting in 1D structure before and after the barriers. 

In order to solve the problem analytically, we first find the well-known transmission $T(E)$ of an incident particle without spin through the double Delta barrier structure with $V_0=0$ in the confined 0D island.  The function $T(E)$ shows resonances corresponding to the confined states, identical for both spin components, because no magnetic fields are acting yet. The maximum of the transmission $T=1$ is obtained as the solution of the equation:

\begin{eqnarray}
4Z^2k^2\frac{\cos(ka)^2}{\sin(ka)^2}+4Z^3k\frac{\cos(ka)}{\sin(ka)}=-Z^4
\label{T1}
\end{eqnarray}\\

Then, the energy difference of the two spin components $V_0$ is taken into account, corresponding to a field along $Z$. The goal is to rotate the in-plane pseudospin by 90 degrees by changing the relative phase of the two spin components at the output barrier.

We find that we can obtain the required $\pi/2$ phase difference between components of spinor wavefunction at the output of the barrier when the value of $V_0$  is fixed at $V_0=\Delta E/2$, where  $\Delta E$ is the energy difference between the two points where $T(E)=1/2$ near a transmission resonance peak. These points are defined by the equation:

\begin{eqnarray}
4Z^2k^2(\cos(2ka)+1)+4Z^3k\sin(2ka)+\nonumber\\
+Z^4(1-\cos(2ka))=\frac{16k^4}{2}
\label{T12}
\end{eqnarray}

where $Z=2mV_1/\hbar^2$ is a constant describing the system's parameters, and the transmission is a function of the wavevector $k=\sqrt{2mE}/\hbar$. 
This phase difference of $\pi/2$ can be obtained for any localized state at the corresponding transmission energies. One should note that the rotation of pseudospin is inevitably accompanied with the decrease of the transmission, because the energy is moved out of resonance by the internal $Z$ effective field. However, this decrease remains relatively small, and the polariton qubit can be later reamplified in intensity without the loss of coherence by stimulated scattering \cite{AmoPrlRestim}.

Once the most important step of the CNOT gate is at hand, the rest is much simpler to describe. If one fixes the in-plane effective fields, whose control is experimentally difficult, it is easy to choose the injection and detection lengths for the input and output wires respectively, in order to obtain the required 90 degrees rotations around the corresponding fields. This distance is given by $x=\pi\hbar k/2\Omega m$, where we assume that the strength of both in-plane fields is the same: $\Omega_X=\Omega_Y=\Omega$, and the wavevector $k$ corresponds to the resonant transmission through the barriers given by Eq.\ref{T1}.

 \section{Numerical simulations}
 
In order to verify our analytical findings, we have also solved the 1D spinor Schrodinger equation numerically:

\begin{equation}
 i\hbar\frac{\partial\Psi}{\partial t}=\left(-\frac{\hbar^2}{2m}\Delta+V(x)+\mathbf{\Omega}(x)\mathbf{\sigma}-\frac{i\hbar}{2\tau}\right)\Psi+P\left(x,t\right) 
 \end{equation}

 In this equation, as compared to Eq. 2, we have included a realistic spatial potential profile, using rectangular barriers instead of delta functions. All parameters were taken similar to that of a realistic polariton resonant tunneling diode structure \cite{HaiSon}. Another important parameter is the polariton lifetime $\tau$: unlike other qubit implementations, polaritons are not only subject to decoherence, they are also disappearing because of the finite escape rate through the cavity mirrors. The coherence time for polaritons is in general considerably longer than the lifetime, and therefore the quantum computing schemes remain possible, although the signal intensity can significantly decrease during the device operation. On the other hand one can remark that the use of non-radiaitve guided polariton modes \cite{Sanvitto,Solnyshkov2014} can strongly limit these losses, keeping other advantages of the polariton system. 
 
 In our simulations, we consider two distinct cases. In the first case, the polaritons corresponding to the target qubit are injected continuously by the pumping $P$, and propagate through the system, decaying during their propagation. The quantum state is continuously detected at the output. This configuration corresponds to the problem solved analytically in the previous section. It can be used in the proof-of-principle experiment, staying very close to the experimentally available configuration\cite{HaiSon}. This case is described in the first subsection.
 
  In the second case, each qubit is created by a pumping pulse localized in space and time, and then the target qubit propagates through the system and is detected at the output. This last case corresponds to the most realistic description of the device operation, and it will be described in the second subsection.

\subsection{Continuous-Wave operation}
Let us start with the case of the continuous pumping. First, we simulate only the action of the field in the $Z$ direction between the barriers. As shown in figure 1, one expects the pseudospin to rotate in plane by 90 degrees: if one injects the qubit in the state $S_X=1$ at the edge of the left barrier, it should be emitted in the state $S_Y=\pm 1$ at the edge of the right barrier.
 
 \begin{figure}[tbp]
 \includegraphics[scale=0.45]{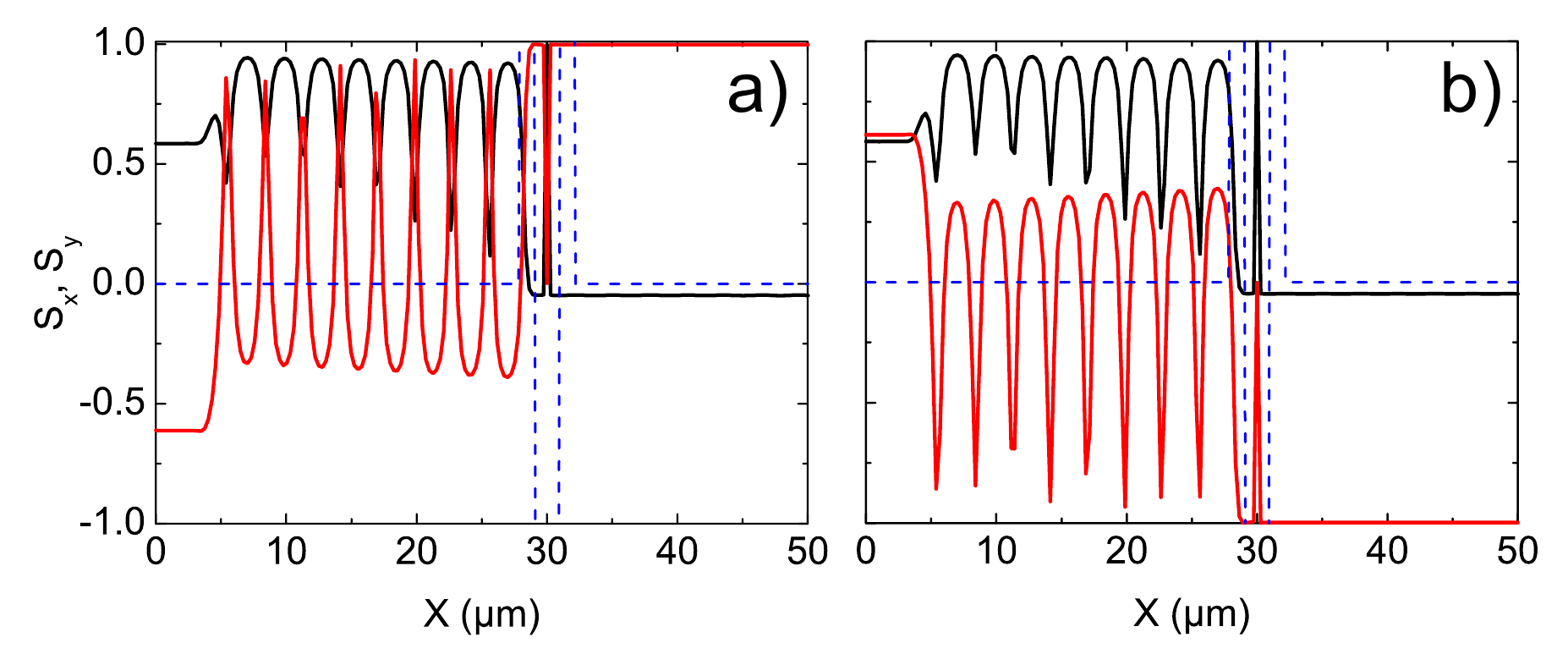}
 \caption{ (color online) In-plane pseudospin rotation due to the field along $Z$ in the trap: the pseudospin components $S_X$ (black) and $S_Y$ (red) as a function of $x$. The blue dashed line shows the potential profile.}
 \label{SxSy}
 \end{figure}
 
Figure \ref{SxSy} shows the results of the simulations: the pseudospin in-plane components plotted in blue and red as a function of coordinates for a steady state regime. The potential barriers (black lines) used in the calculations are 4 meV in height, with a width of 1.37 $\mu$m. Positive effective magnetic field between barriers represents the action of the control qubit in the state $\ket{0}$. At the output of the barriers, $S_Y$=1, as expected for the key element of the CNOT gate (step 2).  The oscillations of the pseudospin component on the left of the double-barrier structure are due to the interference of incident and reflected waves. In numerical simulations, we have taken the injection spot as the $x=0$ reference. In both cases, the absolute value of the $Y$ projection of the pseudospin is very close to 1 (red line either at $+1$ or at $-1$), which demonstrates the high efficiency of the gate operation.

Next, we show the results of the simulation of the whole  CNOT gate with effective fields in the input and output arms. The target qubit is injected in the state $\ket{0}$ and should become either $\ket{0}$ or $\ket{1}$ at the output, depending on the state of the control qubit (sign of the effective field in the $Z$ direction). In order to check this, we plot the $S_Z$ pseudospin component as a function of $x$, which means measuring the circular polarization degree of light from the experimental point of view.

 \begin{figure}[tbp]
 \includegraphics[scale=0.8]{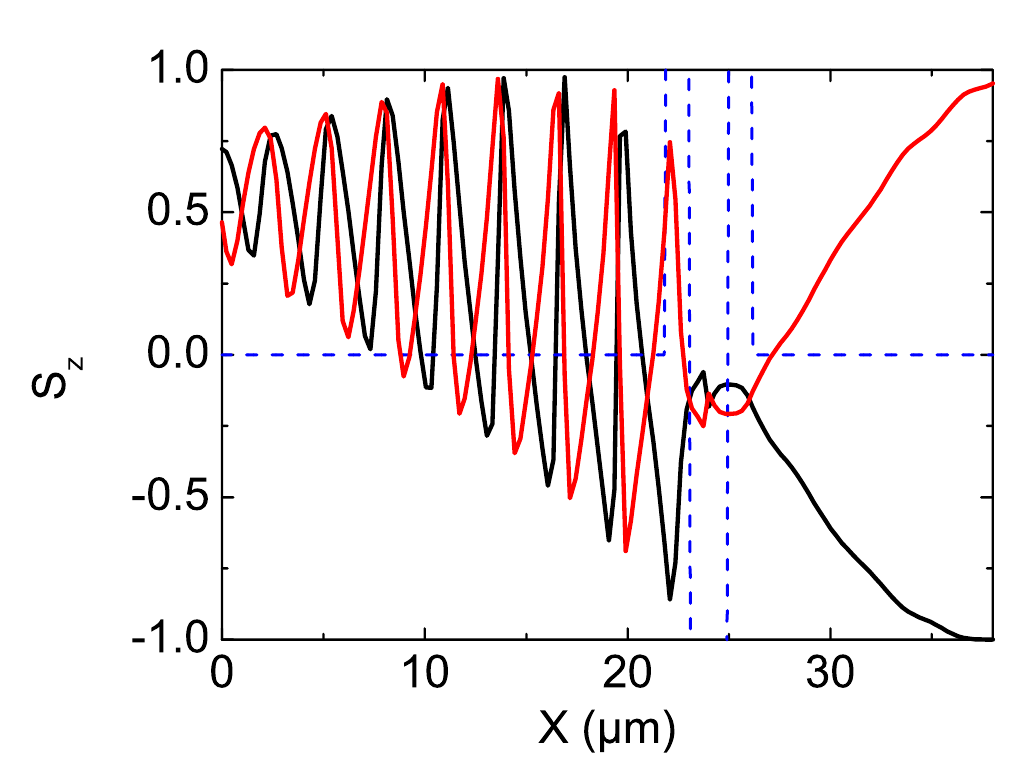}
 \caption{(color online) CNOT gate in \emph{cw} regime: $S_Z$ as a function of $x$ for two cases: black - control qubit is in the state $\ket{1}$, red - control qubit is $\ket{0}$. Blue dashed line indicates the potential profile. }
 \label{Sz}
 \end{figure}

Figure \ref{Sz} demonstrates the results of this simulation for two cases. We can see that for a control qubit in state $\ket{1}$ (negative field, black line), $S_Z$ is equal to -1 at the output of the structure which means that target qubit is converted into the state $\ket{1}$. In the same way, with a control qubit in state $\ket{0}$ (positive field, red line),  the target qubit remains in the state $\ket{0}$. This figure demonstrates the expected operation of the CNOT gate, implemented on a scale of 40 $\mu$m, well below the coherence length for polaritons \cite{Wertz2010}.

\subsection{Pulsed operation}
Finally, we simulate the operation of the device in the pulsed regime. The pulse setting the control qubit is spatially localized on the quantum trap between the barriers and its energy is tuned in resonance with the ground state of the trap. The pulse setting the target qubit is spatially localized at a calculated distance before the first barrier, according to the previous calculations, and its energy is tuned in resonance to the first excited state of the trap, corresponding to its first transmission resonance.

 \begin{figure}[tbp]
 \includegraphics[scale=0.45]{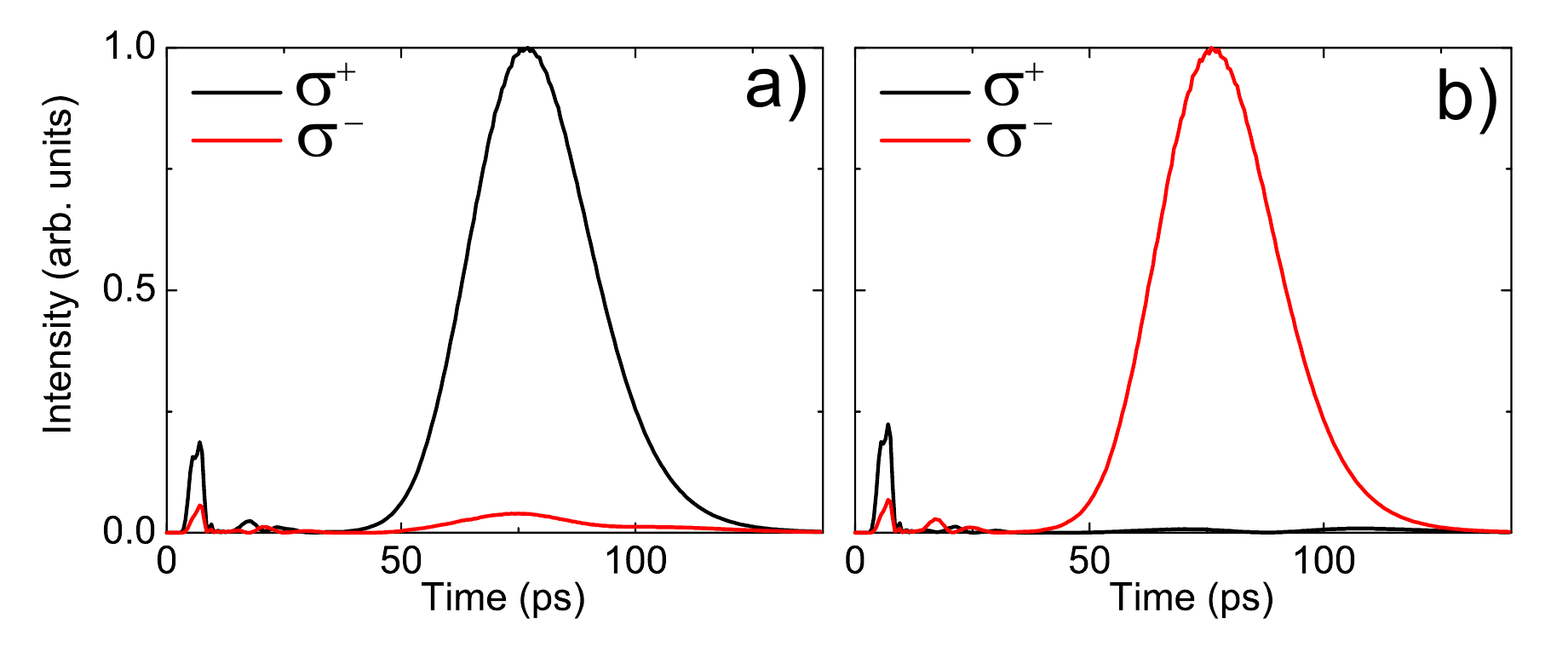}
 \caption{ (Color online) CNOT gate in the pulsed regime: intensities of the circular components at the output of the gate (detection spot). Panel a): control qubit created in the state $\ket{0}$ . Panel b): control qubit created in the state $\ket{1}$}
 \label{pulsed}
 \end{figure}

The results of the simulations are presented in figure 4. We plot the intensity of the two spin components at the output of the device. Panel a) corresponds to control qubit in the state $\left|0\right\rangle$, and panel b) corresponds to the control qubit in the state $\left|1\right\rangle$. The small maximum appearing at around 5~ps corresponds to the particles escaping from the quantum trap at the moment of the injection of the control qubit. The main maximum corresponds to the target qubit passing through the detector. We see that the pulse of the target qubit passes through the device almost undisturbed, and that its polarization is inverted or not, depending on the state of the control qubit. It is precisely what one expects from the CNOT gate.  The duration of the control pulse was 5 ps, and the duration of the target pulse was 20 ps. This duration can be reduced in order to increase the operating frequency of the device, but this will lead to the decrease of its efficiency, decreasing the circular polarization degree of the target qubit at the output. The intensity of the target qubit should be significantly smaller than that of the control qubit, in order to keep the one-way action of the control qubit on the target one, otherwise the state of the control qubit is perturbed after the operation cycle of the gate.

\section{Conclusions}
To conclude, we have proposed an implementation and demonstrated the operation of an all-optical CNOT quantum gate based on a cavity polariton circuit very similar to one recently implemented experimentally \cite{HaiSon}. These results are promising for the scalability of future photonic quantum computing. We can easily add other similar structures in order to implement several gates to build quantum circuitry. The proof-of-principle experiment based on existing structures appears quite feasible for such systems. Similar schemes, but based on guided polariton modes could strongly limit radiative losses and the decay of the signal during the time of flight\cite{Sanvitto}. It could also facilitate the use of large band gap semiconductors, such as GaN or ZnO, which could allow to envisage room temperature operation \cite{Solnyshkov2014}.

\bibliography{biblio}

\end{document}